# Massively parallel dual-comb molecular detection with subharmonic optical parametric oscillators

V. O. Smolski[1,2], H. Yang[1,3], J. Xu[1], and K. L. Vodopyanov[1*]

[1] CREOL, College of Optics and Photonics, Univ. Cent. Florida, Orlando, FL 32816, USA
[2] IPG Photonics-Mid-Infrared Lasers, Birmingham, Alabama 35203, USA
[3] Dept. of Precision Instruments, Tsinghua University, Beijing, 100084, China.
* *vodopyanov@creol.ucf.edu*

**Mid-infrared (mid-IR) spectroscopy offers unparalleled sensitivity for the detection of trace gases, solids and liquids, which is based on the existence of strong telltale vibrational bands in this part of the spectrum. It was shown more than a decade ago that a dual-comb Fourier spectroscopy could provide superior spectral coverage combined with high resolution and extremely fast data acquisition. Capabilities of this method were limited because of difficulty of producing twins of mutually coherent frequency combs in the mid-IR. Here we report a phase-coherent and broadband dual-comb system that is based on a pair of subharmonic (frequency-divide-by-two) optical parametric oscillators, pumped in turn by two phase-locked thulium fiber lasers at $\lambda \approx 2$ μm. We demonstrate simultaneous detection of multiple molecular species in the whole band of 3.2-5.3 μm (frequency span 1200 $cm^{-1}$) augmented by the pump laser band of 1.85-2 μm (span 400 $cm^{-1}$), with spectral resolution 0.01-0.07 $cm^{-1}$ and acquisition speed of up to 160,000 resolution elements in few seconds.**

Coherent laser beams provide a unique prospect for sensing molecules through their resonant absorption features – either remotely or via multipass action. A privileged window for ultrasensitive molecular spectroscopy is the fingerprint region (2.5 to 20 μm) of the mid-IR spectrum, where the strongest absorption lines can be addressed. For example, better than 0.1 part-per-trillion level of sensitivity for molecular detection (that is less than 1 molecule per $10^{13}$ molecules of the buffer gas) has been already demonstrated [1]. Frequency combs – manifolds of evenly spaced and phase locked narrow spectral lines that are produced by phase-stabilized femtosecond lasers – were introduced in the late 1990s and have revolutionized accurate measurements of frequency and time [2]. In the mid-IR range, the advent of optical combs brought a new set of tools that substantially improved both precision and sensitivity of molecular detection [3].

Spectroscopic applications of optical frequency combs include those where (i) a broadband comb directly interrogates an absorbing sample after which the spectrum is dispersed in two dimensions and sensed with detector arrays [4,5,6], (ii) Michelson interferometer based Fourier transform spectroscopy [7-11], and (iii) dual-comb spectroscopy based on superimposing two coherent frequency-comb beams [12-23]. In the latter case, a second frequency comb, with a small offset of mode spacing (pulse repetition frequency) plays a role of the time-delayed second arm in the Michelson interferometer. The benefits include orders of magnitude improvement in acquisition speed over standard Fourier-transform spectroscopy and high-resolution capability. This technique however requires high mutual coherence between the two combs and most reports that take full advantage of the dual comb technique (broad coverage, intermodal resolution, rapid scans) have so far privileged the near-infrared domain [15,16,17].

The true promise of dual-comb spectroscopy lies in the mid-IR domain and proof-of-principle demonstrations have been carried out in the 10-μm wavelength region (instantaneous spectral span 250 cm$^{-1}$, spectral resolution 2 cm$^{-1}$) [13], in the 2.4-μm region (span 200 cm$^{-1}$, resolution 2 cm$^{-1}$)[18], and in the 3-μm region (150 cm$^{-1}$ / 0.2 cm$^{-1}$ [19]; 250 cm$^{-1}$ / 0.8 cm$^{-1}$ [20]; and 350 cm$^{-1}$ / 0.2 cm$^{-1}$ [21]). Also, with narrower spectral coverage, high resolution has been demonstrated in the 3.4 -μm region (span 30 cm$^{-1}$; resolution 10 kHz ≈3x10$^{-7}$ cm$^{-1}$) [22], and in the 7-μm region (16 cm$^{-1}$ / 0.003 cm$^{-1}$) [23].

For simultaneous detection of an assortment of molecules with different functional groups, one needs a spectrally broad comb, ideally spanning an octave or more and with spectral resolution better than 0.1 cm$^{-1}$. Here we report a dual-comb spectrometer that uses a pair of broadband (3.2 - 5.3 μm) phase-coherent subharmonic OPOs and demonstrate massively parallel detection of molecular species with up to 160,000 resolution elements in few seconds recording time.

It has been demonstrated recently that a femtosecond optical parametric oscillator (OPO) operating at degeneracy, while acting as nearly perfect optical frequency divider, dramatically augments the pump spectrum [24,25,26] and, most important, preserves phase coherence of the pump [27]. For example, the relative linewidth between a degenerate OPO and a pump frequency comb was shown to be well below 1 Hz [28]. A highly coherent frequency comb with record-wide instantaneous wavelength span of 2.6-7.5 μm was achieved in a subharmonic GaAs-based OPO pumped by a Tm-fiber laser [29].

Our dual-comb mid-IR system starts with a pair of phase-coherent optically referenced Tm-fiber-laser frequency combs [30,31] serving as pump with the following parameters: central wavelength 1.93 μm, repetition rate 115 MHz, pulse duration 90 fs, and average power ~300 mW for each laser. For Tm-fiber comb stabilization, an octave-spanning supercontinuum (SC) was generated from each laser, and used to phase lock, via $f$-to-$2f$ interferometry [2], carrier envelope offset (CEO) frequency to the same value of $f_0 \approx$ 40 MHz for both lasers. Phase locking one of comb teeth relative to a common narrow-linewidth diode-laser optical reference at 1563.88 nm (linewidth ~3 kHz, long-term drift ~100 kHz) was used to stabilize the repetition rate $f_{rep}$ for both lasers. The repetition-rate offset between the two lasers was kept at $\Delta f_{rep} \approx$ 69.26 Hz, such that a dual-comb interferogram appeared every ~14 ms. Since the same optical reference was shared between the two Tm-fiber lasers, we observed a very high stability (± 10 μHz) of the repetition-rate offset.

The twin Tm-laser system pumped two identical subharmonic ring-cavity OPOs, based on 0.5-mm-long orientation-patterned GaAs (OP-GaAs) crystals as gain elements [29] (Fig. 1). The OPO cavities were purged with dry air and the pumping threshold was typically 10 mW, in terms of the average pump power. The OPOs were actively stabilized for the doubly-resonant degenerate operation using a dither-and-lock method [25] and were running in a strictly degenerate mode such that the CEO frequency for each OPO was half of that of the pump: $f_0/2 \approx$ 20 MHz. This was confirmed by monitoring radio frequency (RF) multiheterodyne beats obtained between SC originating from the pump laser and parasitic sum frequency (pump + OPO) light originating from the OPO cavity, as described in [29]. Each OPO shared coherence properties of the pump, with an estimated sub-Hz relative linewidth. For both OPOs the instantaneous spectral span was 3.2-5.3 µm at -20dB level and the output power was kept at 20 mW.

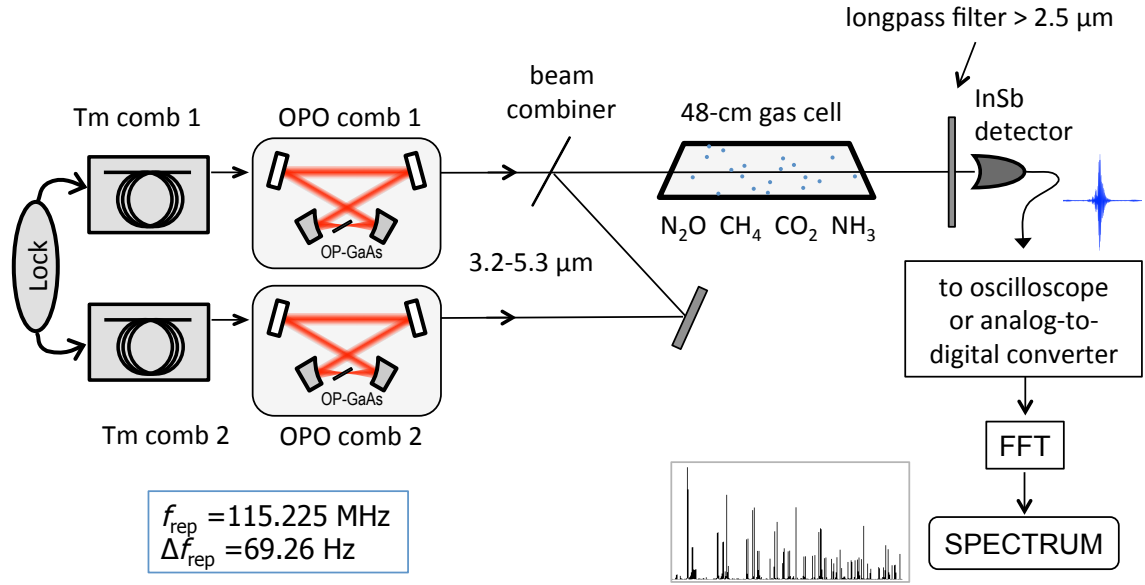

**Fig. 1** Dual-comb spectrometer based on subharmonic OPOs. A pair of phase-locked Tm-doped fiber lasers ($\lambda \approx 2$ µm, $f_{rep}$ =115 MHz, $\Delta f_{rep}$ =69 Hz) was used to pump a twin OPO system. The OPO output beams were combined, passed through a 48-cm-long gas cell, detected with an InSb detector, digitized, and Fourier transformed to retrieve an optical spectrum. A longpass (> 2.5 µm) filter was used to cut the pump radiation.

The two OPO beams were spatially combined, passed through a 48-cm-long gas cell containing a mixture of molecular gases, and sent to an InSb detector (77K, 60 MHz, long-wave cut-off at 5.6 µm). In the first trial the two gases, nitrous oxide ($N_2O$) at 186 parts-per-million (ppm) and methane ($CH_4$) at 1250 ppm concentration were mixed in the optical cell with the buffer nitrogen ($N_2$) gas at $p$=1 atm pressure. We coherently averaged 500 interferograms (total measurement time 7 sec) from the detector output using a digital oscilloscope with 1M sample record length and only 8-bit vertical resolution. The resulting time-domain interferogram is shown in Fig. 2(a).

The recording window in this experiment was 800 µs, limited by the memory of the oscilloscope. By taking a Fast Fourier Transform (FFT) of the interferogram, we obtained a radio frequency (RF) spectrum that spanned from 35 to 55 MHz and is shown in Fig. 2(b). The optical spectrum on the same figure was reconstructed by up-scaling the frequency axis by a factor $f_{rep}/\Delta f_{rep}$, according to the dual-comb spectroscopy method [13]. One can see a broad OPO spectrum modulated by molecular absorption. By zooming into different portions of the 1240 $cm^{-1}$-wide spectrum we were able to observe absorption features of different molecules recorded *simultaneously*. The optical resolution of 0.07$cm^{-1}$, determined by the 800-µs recording window, was adequate in this experiment to fully resolve ro-vibrational lines of several molecules with a typical linewidth in the range 0.15-0.20 $cm^{-1}$ at $p$=1 atm.

Figs. 3(a,b) depict normalized spectra (we used polynomial baseline fitting) of $N_2O$ and $CH_4$ and show good agreement with the HITRAN database simulation [32]. Moreover, the optical spectra

that were obtained by referencing to radio frequencies only show better than 0.01 cm$^{-1}$ absolute frequency accuracy. Also, Figs. 3(c,d) show measured absorption features of several molecules naturally present in the atmosphere: isotopic carbon dioxide ($^{13}CO_2$) at nominal 4 ppm concentration, water at estimated ~ 0.1% concentration in the purged OPO cavity, and methane at ~ 1.7 ppm natural concentration. Detection sensitivity in the case of Fig. 3(c,d) was enhanced by an intracavity effect described in [8], since these molecules were present *inside* the OPO cavity (we estimate the effective propagation length to be 20-30 m, due to a finite lifetime of a photon in the resonator). We should note that the intracavity action accounts for the asymmetry (dispersive feature) of $^{13}CO_2$ absorption lines in Fig. 3(c), since molecular resonances affect, via Kramers-Kronig relation, group dispersion inside the cavity [8].

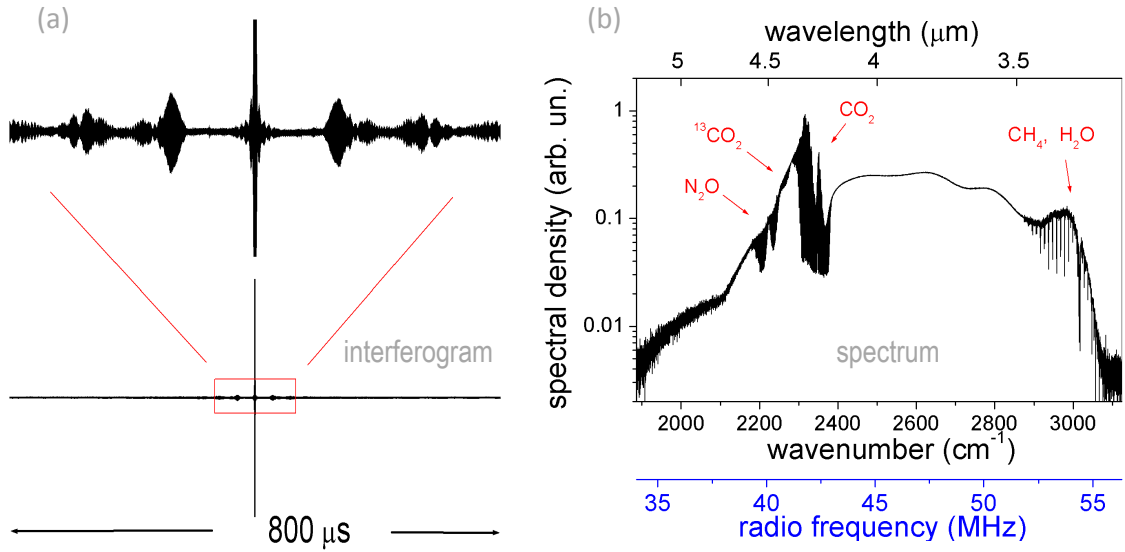

**Fig. 2** Dual-comb interferogram and retrieved RF/optical spectrum. (a) The original interferogram obtained from the InSb detector and (b) RF spectrum obtained via Fast Fourier Transform. The optical spectrum was retrieved via up-scaling the frequency axis by $f_{rep}/\Delta f_{rep}$. Absorption features due to different molecules (N$_2$O, CH$_4$, CO$_2$, isotopic CO$_2$, and H$_2$O) are indicated by arrows.

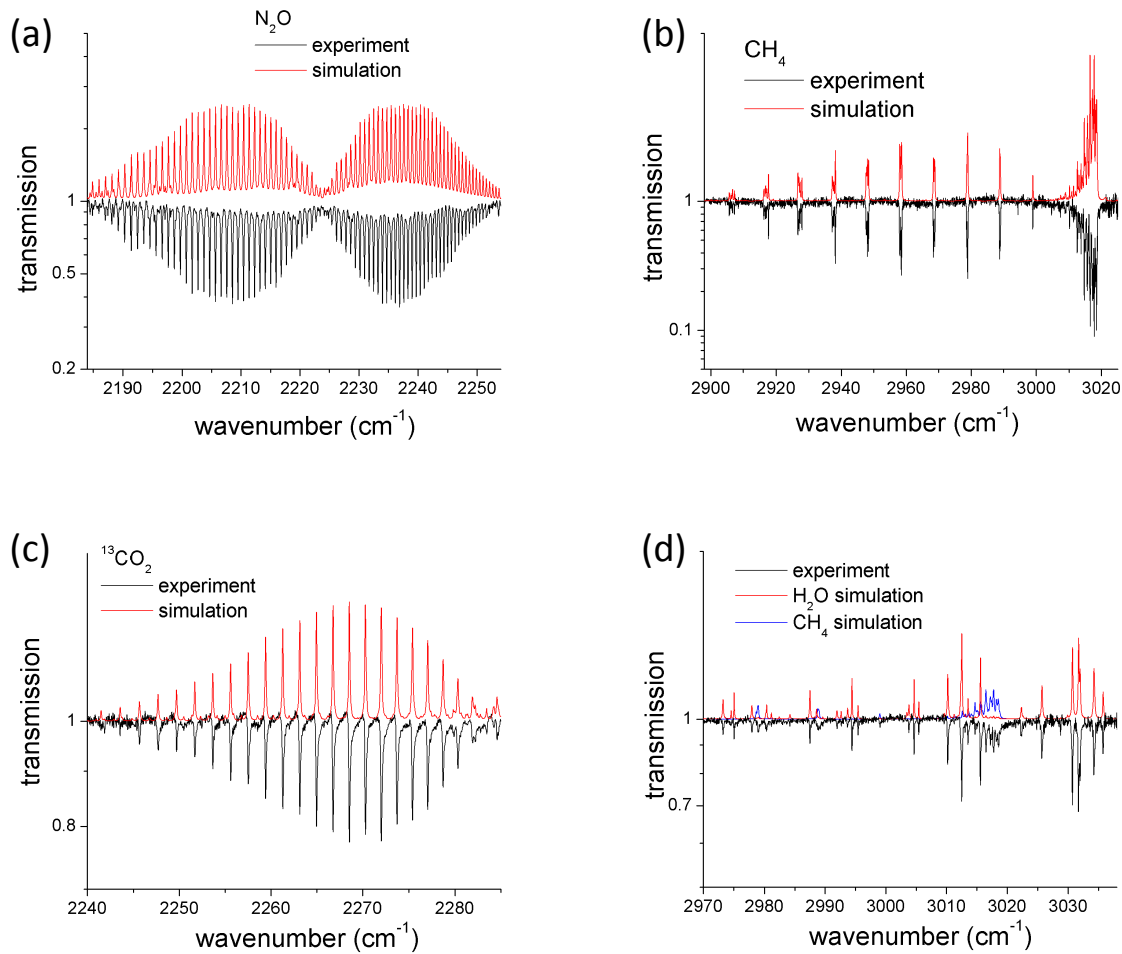

**Fig. 3.** Dual-comb mid-IR spectra. Normalized transmission of a 48-cm gas cell for (a) nitrous oxide ($N_2O$) (186 ppm in 1-atm $N_2$ buffer gas) and (b) methane ($CH_4$) (1250 ppm in 1-atm $N_2$ buffer gas). Also shown are spectra for: (c) isotopic carbon dioxide ($^{13}CO_2$) and (d) water ($H_2O$) and methane ($CH_4$) – all present in the atmosphere at trace amounts. The theoretical spectra are inverted for clarity.

By removing the longpass filter before InSb detector in Fig. 1(a) we were able to recover from the dual-comb interferogram both the OPO spectrum and the pump spectrum at 1.85-2 µm, as shown in Fig. 4(a). To demonstrate high spectral resolution of our method, we filled the optical cell with gases at reduced pressure, such as $CO_2$ (at 11% concentration with air at $p$=0.5 atm as buffer gas) and $NH_3$ (at 0.6% concentration with air at 0.15 atm as buffer gas). In this experiment, we used a more advanced detection method with a 16-bit 250 MS/s analogue-to-digital card. This allowed increasing the recording time window to 5.2 ms corresponding to the optical resolution of 0.01 cm$^{-1}$. We coherently averaged 300 interferograms with the measurement time ~4 sec.

Figs. 4(b,c) display normalized transmission spectra for $CO_2$ and $NH_3$. The measured linewidth for $CO_2$ (0.069 cm$^{-1}$) and for $NH_3$ (0.028 cm$^{-1}$) are both in good agreement with the HITRAN database and are consistent with our expected (0.01 cm$^{-1}$) spectral resolution.

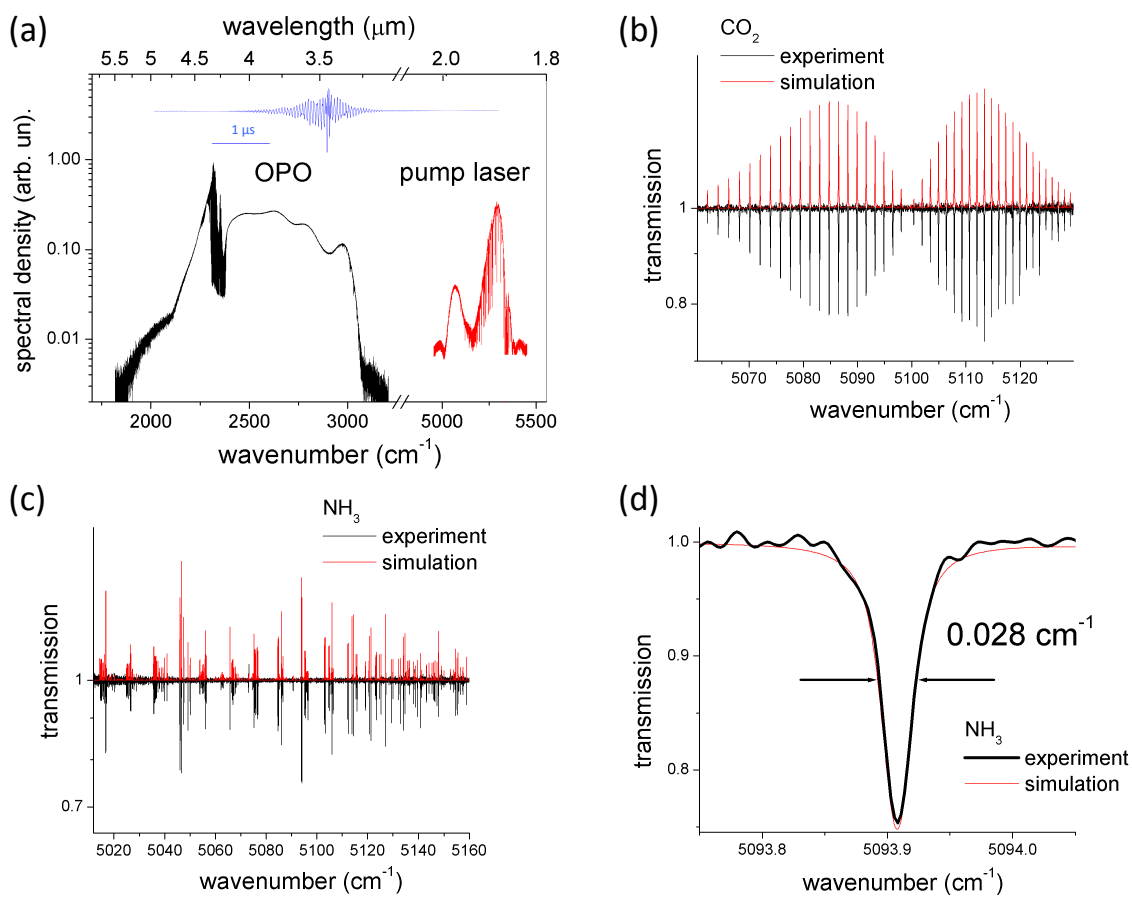

**Fig. 4** High-resolution spectral measurements in the 1.85 - 2 μm spectral band. (a) Compound (OPO + pump) optical spectrum retrieved from the interferogram, central portion of which is shown on top. (b) Normalized transmission spectra for carbon dioxide ($CO_2$) (11% in 0.5-atm air) and (c) ammonia ($NH_3$) (0.6% in 0.15-atm air) (d) Expanded view of the $NH_3$ line shape. The theoretical spectra are inverted for clarity in (a-c).

In conclusion, we present our results on dual-comb mid-IR spectroscopy exploiting two low-threshold and broadband subharmonic OPOs with high mutual coherence. The instantaneous detection bandwidth was 3.2-5.3 μm (1240 cm$^{-1}$) augmented by the Tm-laser band 1.85-2.0 μm (400 cm$^{-1}$), and spectral resolution 0.01-007 cm$^{-1}$, which corresponds to up to 160,000 resolution elements obtained in a few seconds of averaging time. The broad instantaneous bandwidth allowed simultaneous detection of five molecules, while high mutual coherence – for averaging of the interferogram without phase correction. Although it was not our goal to demonstrate the ultimate detection sensitivity, we believe that sensitivities of < 1 part-per-billion (ppb) are achievable with our dual-comb technique using multipass, external cavity or intracavity detection methods, where ppb-level sensitivity has been already demonstrated with single frequency combs [7,8,9,10].

Acknowledgements.

Research was conducted at the College of Optics and Photonics (CREOL) at Univ. Central Florida. Support was provided by the Office of Naval Research (ONR), grant number N00014-15-1-2659; Defense Advanced Research Projects Agency (DARPA) SCOUT program through grant number W31P4Q-15-1-0008 from AMRDEC. We thank Kevin Lee and Jie Jiang for

References

1. I. Galli, S. Bartalini, S. Borri, P. Cancio, D. Mazzotti, P. De Natale, and G. Giusfredi, Molecular Gas Sensing Below Parts Per Trillion: Radiocarbon-Dioxide Optical Detection, Phys. Rev. Lett. 107, 270802 (2011).

2. T. Udem,  R. Holzwarth, and T.W. Hänsch, Optical frequency metrology, Nature 416, 233-237 (2002).

3. A. Schliesser, N. Picqué, and T. W. Hänsch, Mid-infrared frequency combs, Nature Photon. 6, 440-449 (2012).

4. S. A. Diddams, L. Hollberg, and V. Mbele, "Molecular fingerprinting with the resolved modes of a femtosecond laser frequency comb," Nature 445, 627–630 (2007).

5. L. Nugent-Glandorf, T. Neely, F. Adler, A. J. Fleisher, K. C. Cossel, B. Bjork, T. Dinneen, J. Ye, and S. A. Diddams, Mid-infrared virtually imaged phased array spectrometer for rapid and broadband trace gas detection, Opt. Lett. 37, 3285-87 (2012).

6. A. J. Fleisher, B. J. Bjork, T. Q. Bui, K. C. Cossel, M. Okumura, and J. Ye, Mid-infrared time-resolved frequency comb spectroscopy of transient free radicals, J. Phys. Chem. Lett. 5, 2241-46 (2015).

7. A. Foltynowicz, P. Mas1owski, T. Ban, F. Adler, K. C. Cossel, T. C. Briles and J. Ye, Optical frequency comb spectroscopy. Faraday Discuss. 150, 23–31 (2011).

8. M. W. Haakestad, T. P. Lamour, N. Leindecker, A. Marandi, and K. L. Vodopyanov, Intracavity trace molecular detection with a broadband mid-IR frequency comb source , J. Opt. Soc. Am. B 30, 631 (2013).

9. S. A. Meek, A. Poisson, G. Guelachvili, T. W. Hänsch, and N. Picqué, "Fourier transform spectroscopy around 3 μm with a broad difference frequency comb," Appl. Phys. B 114(4), 573–578 (2014).

10. A. Khodabakhsh, V. Ramaiah-Badarla, L. Rutkowski, A. C. Johansson, K. F. Lee, J. Jiang, C. Mohr, M. E. Fermann, and A. Foltynowicz, Fourier transform and Vernier spectroscopy using an optical frequency comb at 3–5.4 μm, Opt. Lett. 41, 2541-2544 (2016).

11. P. Maslowski, K. F. Lee, A. C. Johansson, A. Khodabakhsh, G. Kowzan, L. Rutkowski, A. A. Mills, C. Mohr, J. Jiang, M. E. Fermann, and A. Foltynowicz, Surpassing the path-limited resolution of Fourier-transform spectrometry with frequency combs, Phys. Rev. A 93, 021802(R) (2016).

12. F. Keilmann,  C. Gohle, and R. Holzwarth, Time-domain mid-infrared frequency-comb spectrometer. Opt Lett 29, 1542-1544 (2004).

13. A. Schliesser, M. Brehm, F. Keilmann, and D. W. van der Weide, Frequency-comb infrared spectrometer for rapid, remote chemical sensing, Opt. Express 13, 9029-9038 (2005).

14. I. Coddington, N. Newbury, and W. Swann, Dual-comb spectroscopy, Optica 3, 414-426 (2016).

15. Coddington, I., Swann, W.C. & Newbury, N.R. Coherent multiheterodyne spectroscopy using stabilized optical frequency combs. Phys. Rev. Lett 100, 013902 (2008).

16. I. Coddington, W. C. Swann, and N. R. Newbury, Time-domain spectroscopy of molecular free-induction decay in the infrared, Opt. Lett. 35, 1395-97 (2010)

17. A. M. Zolot, F. R. Giorgetta, E. Baumann, J. W. Nicholson, W. C. Swann, I. Coddington, and N. R. Newbury, Direct-comb molecular spectroscopy with accurate, resolved comb teeth over 43 THz, Opt. Lett. 37, 638-640 (2012).

18. B. Bernhardt, E. Sorokin, P. Jacquet, R. Thon, T. Becker, I.T. Sorokina, N. Picqué, T.W. Hänsch, Mid-infrared dual-comb spectroscopy with 2.4 μm $Cr^{2+}$:ZnSe femtosecond lasers, Appl. Phys. B 100, 3–8 (2010).

19. Z. Zhang, T. Gardiner, and D. T. Reid, Mid-infrared dual-comb spectroscopy with an optical parametric oscillator, Opt. Lett., 38, 3148 (2013).

20. F. C. Cruz, D. L. Maser, T. Johnson, G. Ycas, A. Klose, F. R. Giorgetta, I. Coddington, and S. A. Diddams, Mid-infrared optical frequency combs based on difference frequency generation for molecular spectroscopy, Opt. Express 23, 26814 (2015).

21. Y.W. Jin, S.M. Cristescu, F.J.M. Harren, J. Mandon, Femtosecond optical parametric oscillators toward real-time dual-comb spectroscopy, Appl. Phys. B 119, 65 (2015).

22. E. Baumann, F. R. Giorgetta, W. C. Swann, A. M. Zolot, I. Coddington, and N. R. Newbury, Spectroscopy of the methane ν3 band with an accurate midinfrared coherent dual-comb spectrometer, Phys. Rev. A 84, 062513 (2011).

23. G. Villares, A. Hugi, S. Blaser, and J. Faist, Dual-comb spectroscopy based on quantum-cascade-laser frequency combs, Nature Commun. 5, 5192 (2014).

24. S. T. Wong, K.L. Vodopyanov, and R.L. Byer, Self-phase-locked divide-by-2 optical parametric oscillator as a broadband frequency comb source, J. Opt. Soc. Am. B 27, 876–882 (2010).

25. N. Leindecker, A. Marandi, R.L. Byer, and K.L. Vodopyanov, Broadband degenerate OPO for mid-infrared frequency comb generation, Opt. Express 19, 6304–6310 (2011).

26. N. Leindecker, A. Marandi, R. L. Byer, K. L. Vodopyanov, J. Jiang, I. Hartl, M. Fermann, and P. G. Schunemann, Octave-spanning ultrafast OPO with 2.6–6.1 μm instantaneous bandwidth pumped by femtosecond Tm-fiber laser, Opt. Express 20, 7046–7053 (2012).

27. A. Marandi, N. Leindecker, V. Pervak, R.L. Byer, K. L. Vodopyanov, Coherence properties of a broadband femtosecond mid-IR optical parametric oscillator operating at degeneracy, Opt. Express 20, 7255 (2012).

28. K. F. Lee, C. Mohr, J. Jiang, P. G. Schunemann, K. L. Vodopyanov, and M. E. Fermann, Midinfrared frequency comb from self-stable degenerate GaAs optical parametric oscillator, Opt. Express 23, 26596-26603 (2015).

29. V. O. Smolski, H. Yang, S. D. Gorelov, P. G. Schunemann, and K. L. Vodopyanov, Coherence properties of a 2.6 -7.5-µm frequency comb produced as subharmonic of a Tm-fiber laser, Opt. Lett. 41, 1388-1391 (2016).

30. J. Bethge, J. Jiang, C. Mohr, M. Fermann, and I. Hartl, Optically referenced Tm-fiber-laser frequency comb, Paper AT5A.3 in Adv. Solid-State Photon. (OSA, 2012).

31. M. E. Fermann and I. Hartl, Ultrafast fibre lasers, Nature Photon. 7, 868–874 (2013).

32. HITRAN on the Web; http://hitran.iao.ru/